\documentclass[sigconf]{acmart}
\AtBeginDocument{%
  }

\usepackage{algorithm}
\usepackage{algpseudocode}
\usepackage{amsmath}
\setcopyright{acmlicensed}
\copyrightyear{2018}
\acmYear{2025}
\acmConference[CHI WS40]{Emerging Practices in Participatory AI Design in Public Sector Innovation}{April 27, 2025}{Yokohama, Japan}
\begin{document}

\title{Democratizing Differential Privacy: A Participatory AI Framework for Public Decision-Making}

\author{Wenjun Yang}
\orcid{0009-0005-0629-9335}
\affiliation{%
  \institution{University of Washington Tacoma}
  \city{Tacoma}
  \state{Washington}
  \country{USA}
}
\email{wy927@uw.edu}

\authornote{Presented at CHI 2025 Workshop WS40: Participatory AI Design in Public Sector Innovation (non-archival). \url{https://participatoryaidesign.github.io/}}

\author{Eyhab Al-Masri}
\affiliation{%
  \institution{University of Washington Tacoma}
  \city{Tacoma}
  \state{Washington}
  \country{USA}
}
\email{ealmasri@uw.edu}

\begin{abstract}
This paper introduces a conversational interface system that enables participatory design of differentially private AI systems in public sector applications. Addressing the challenge of balancing mathematical privacy guarantees with democratic accountability, we propose three key contributions: (1) an adaptive $\epsilon$-selection protocol leveraging TOPSIS multi-criteria decision analysis to align citizen preferences with differential privacy (DP) parameters, (2) an explainable noise-injection framework featuring real-time Mean Absolute Error (MAE) visualizations and GPT-4-powered impact analysis, and (3) an integrated legal-compliance mechanism that dynamically modulates privacy budgets based on evolving regulatory constraints. Our results advance participatory AI practices by demonstrating how conversational interfaces can enhance public engagement in algorithmic privacy mechanisms, ensuring that privacy-preserving AI in public sector governance remains both mathematically robust and democratically accountable.
\end{abstract}


\settopmatter{printacmref=false}

\maketitle

\textbf{Reference Format}\\
Wenjun Yang and Eyhab Al-Masri. 2025. \textit{Democratizing Differential Privacy: A Participatory AI Framework for Public Decision-Making.} \\
Presented at CHI 2025 Workshop WS40: Emerging Practices in Participatory AI Design in Public Sector Innovation (non-archival workshop), Yokohama, Japan.

\section{Introduction}
Public sector organizations face a critical challenge in adopting AI systems that balance statistical utility with provable privacy guarantees. As governments deploy AI-driven decision-making tools for public services, urban planning, and resource allocation, ensuring privacy protection while maintaining public trust and transparency is paramount \cite{gritzalis2017transparency}.

Differential Privacy (DP) provides a mathematically rigorous framework for privacy-preserving analytics \cite{Dwork2006}, preventing individual identification through controlled noise injection. However, its adoption in public sector applications is hindered by complex trade-offs involving:

\begin{itemize}
\item \textbf{$\epsilon$-Selection}: Balancing privacy protection with data utility.
\item \textbf{Data Sensitivity}: Adjusting noise for different types of public data (e.g., census, health, mobility).
\item \textbf{Public Accountability}: Ensuring privacy decisions reflect democratic values and remain transparent.
\end{itemize}

Existing approaches fail to democratize privacy decisions in public AI systems. Traditional methods either rely on expert-defined DP settings \cite{Zhang2020} or oversimplify privacy controls through binary opt-in/out interfaces \cite{Lau2018}, excluding meaningful public participation and eroding trust.

To address these challenges, we propose a participatory AI approach that integrates civic engagement into DP decision-making via a conversational interface. This system enables stakeholders, including policymakers and the public, to explore and influence privacy configurations in real time, shifting privacy decisions from top-down expert control to democratic deliberation.

Our system introduces three key innovations: (1) an adaptive $\epsilon$-selection mechanism leveraging TOPSIS-based multi-criteria decision analysis (MCDA) \cite{TOPSIS} to align privacy settings with public priorities; (2) explainable noise injection with real-time Mean Absolute Error (MAE) visualizations and GPT-4-powered impact analysis to enhance transparency and trust; and (3) dynamic legal-compliance constraints that adjust privacy budgets to evolving regulations.

By embedding participatory mechanisms into DP decision-making, our work operationalizes democratic values in privacy governance. This contributes to broader efforts to develop accountable, community-driven AI in public sector innovation, bridging the gap between technical privacy guarantees and citizen participation.

\section{Related Work}

Recent research in human-computer interaction (HCI) and AI governance has emphasized co-design approaches in public sector AI, advocating for participatory frameworks that enhance citizen engagement \cite{Muller2021}. However, technical privacy mechanisms—particularly Differential Privacy (DP)—remain largely opaque to non-expert stakeholders. The mathematical complexity of DP and the lack of intuitive interfaces create a disconnect between privacy-preserving AI techniques and democratic governance.

Zhang et al. \cite{Zhang2020} identify three key barriers to DP adoption in civic contexts: (1) \textit{mathematical complexity}, which makes it difficult for policymakers and citizens to understand privacy protections; (2) \textit{opaque trade-offs}, where the impact of privacy budgets ($\epsilon$) on data utility is unclear; and (3) \textit{lack of stakeholder input channels}, preventing meaningful civic participation in privacy configurations. These barriers often result in top-down, expert-driven privacy decisions that exclude affected communities.

To improve transparency, prior work has explored explainable privacy techniques that clarify DP mechanisms \cite{Lau2018}. However, most approaches rely on static visualizations rather than interactive tools that enable stakeholders to actively engage in privacy decisions.

Chatbots in government services typically support informational queries \cite{Clarke2021} but rarely facilitate algorithmic co-design. Our system advances civic interaction paradigms by implementing a stateful dialogue manager that (1) tracks privacy budget allocations, (2) maintains versioned dataset states, and (3) enables collaborative $\epsilon$-tuning through natural language and visual sliders. This hybrid interface addresses gaps in participatory AI toolkits \cite{Muller2021} by integrating symbolic parameter controls with neural language explanations, enhancing transparency and stakeholder engagement.

We propose a participatory DP framework that enhances democratic engagement via a conversational interface. Using a TOPSIS-based MCDA model for $\epsilon$-selection, it improves privacy trade-off interpretability through (a) \textit{constrained parameter spaces}, (b) \textit{visual decision matrices}, and (c) \textit{interactive weight sliders}, embedding transparency and accountability in public sector AI.
\section{System Design}
\begin{figure}[h]
\centering
\includegraphics[width=0.9\columnwidth]{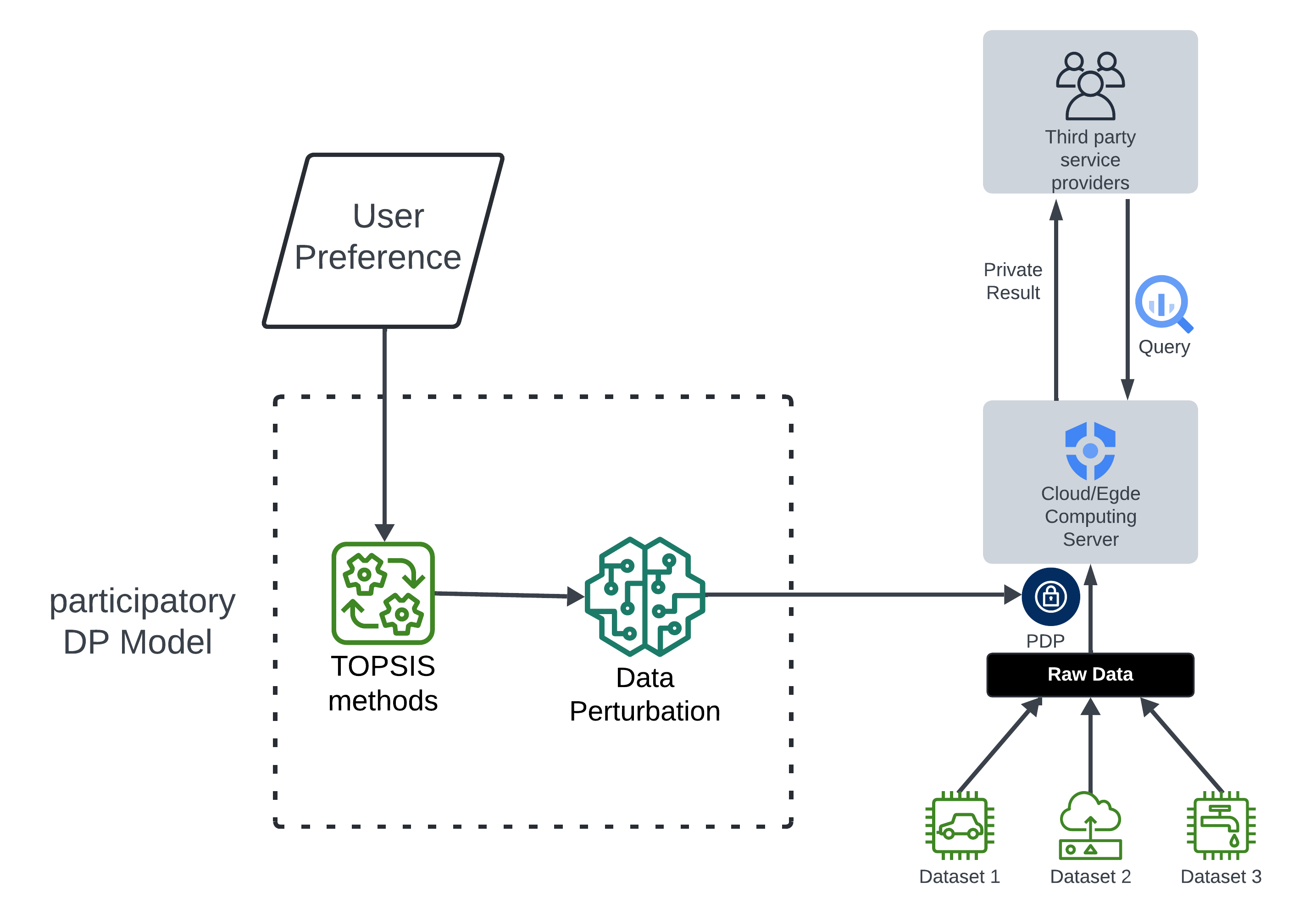}
\caption{System architecture showing the participatory DP workflow}
\label{fig:arch}
\end{figure}

Our participatory DP system integrates web-based interaction with algorithmic privacy controls (Fig. \ref{fig:arch}). The Flask backend consists of three core components:

\begin{itemize}
\item \textbf{Preference Elicitation}: Users specify priorities via sliders for privacy (1-5), accuracy (1-5), legal compliance (yes/no), and data sensitivity (1-3).
\item \textbf{Adaptive $\epsilon$ Selection}: Implements TOPSIS multi-criteria decision analysis \cite{TOPSIS} to resolve trade-offs:
\begin{equation}
\epsilon^* = \underset{\epsilon \in \{0.1,0.5,1.0,1.5,2.0\}}{\arg\max} \frac{D^-}{D^+ + D^-}
\end{equation}
where $D^+$/$D^-$ denote distances to ideal/anti-ideal solutions.
\item \textbf{Conversational Analysis}: Uses GPT-4 to generate natural language explanations of DP impacts.
\end{itemize}

The participatory configuration framework for differential privacy (Algorithm~\ref{alg:dp_config}) enables users to balance privacy-utility trade-offs through an interactive process. By translating user priorities for privacy, accuracy, and regulatory compliance into normalized mathematical weights, the system automates the selection of an optimal privacy budget $\epsilon$  using multi-criteria decision analysis. 

\begin{algorithm}
\caption{Participatory DP Configuration}
\label{alg:dp_config}  
\begin{algorithmic}
\State 1. User uploads dataset and sets privacy, accuracy, and compliance priorities.
\State 2. Normalize slider inputs to compute weights $w_i$.
\State 3. Construct decision matrix for $\epsilon$ alternatives.
\State 4. Compute TOPSIS scores and select optimal $\epsilon^*$.
\State 5. Apply Laplace noise $\mathcal{N} \sim \text{Lap}(\Delta f/\epsilon^*)$.
\State 6. Generate MAE visualization and GPT-4 impact analysis.
\State 7. Present interactive report with refinement suggestions.
\end{algorithmic}
\end{algorithm}

\section{Evaluation}
\subsection{Experimental Validation of Participatory DP}
We evaluated our framework using computational simulations on the Household Electricity Demand (HED) dataset \cite{muratori2017impact}, which provides power consumption profiles for 200 randomly selected households from the 2009 Midwest RECS dataset. The dataset captures realistic residential electricity usage patterns, validated against metered data. Each profile records power consumption (in watts) at a 10-minute resolution, accounting for household size and occupancy variations.

\begin{table}[h]
\caption{Impact of User Preferences on DP Outcomes}
\label{tab:simresults}
\centering
\begin{tabular}{lccc}
\textbf{Metric} & \textbf{Privacy-First} & \textbf{Balanced} & \textbf{Utility-First} \\ \hline
Selected $\epsilon$ & 0.1 & 1.0 & 2.0 \\
MAE (kWh) & 83.2 & 9.6 & 3.3 \\
Privacy Score\textsuperscript{*} & 4.8 & 3.2 & 2.1 \\
\end{tabular}
\vspace{2mm}
\footnotesize{\textsuperscript{*}GPT-4 generated privacy ratings on a 1-5 scale.}
\end{table}

Our results confirm the expected trade-off between privacy and accuracy. The strong negative correlation between $\epsilon$ and MAE ($r = -0.96$, $p < 0.01$) aligns with DP principles, demonstrating that privacy-first configurations introduce 3.6$\times$ more noise than utility-optimized settings. This validates our approach’s ability to dynamically balance privacy and utility based on user-defined preferences.

\begin{figure}[h]
\centering
\includegraphics[width=1\columnwidth]{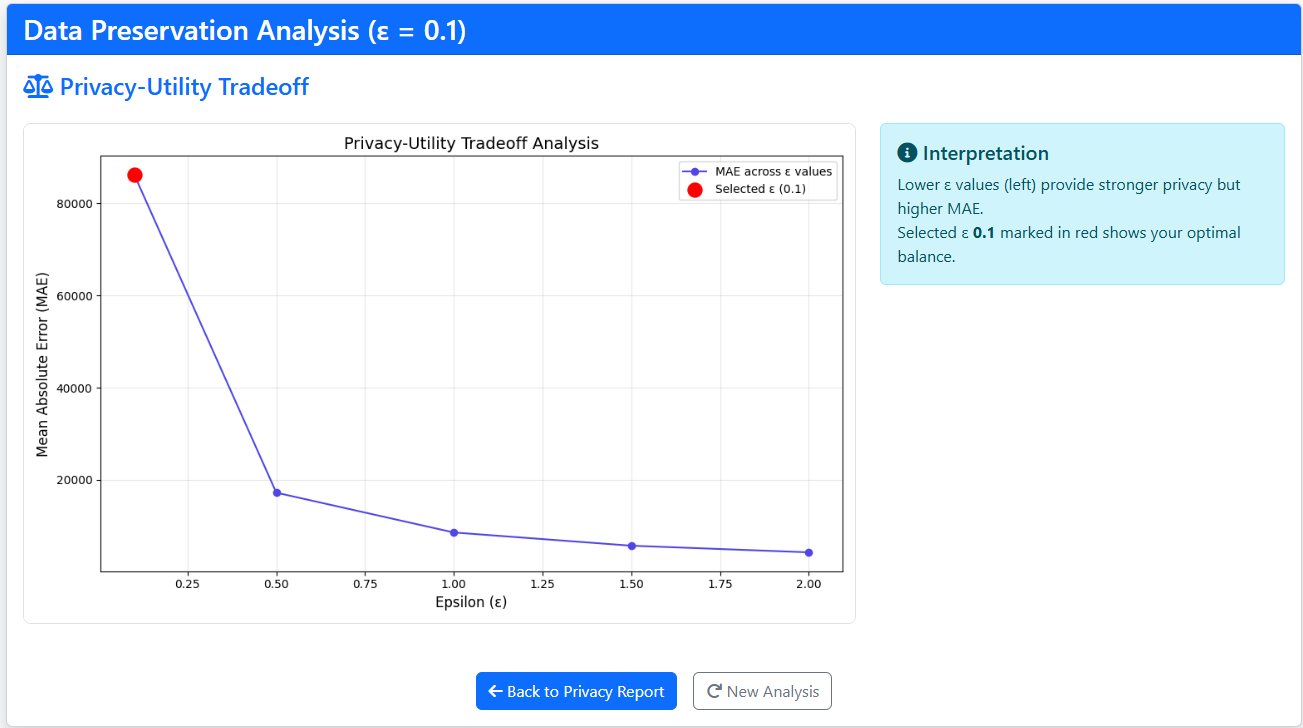}
\caption{Simulated $\epsilon$ selection under different preference profiles}
\label{fig:tradeoffs}
\end{figure}

\subsection{Privacy-Utility Trade-off Analysis}
The results in Fig. \ref{fig:tradeoffs} highlight the significant impact of DP on data utility, where injected noise disrupts time series patterns. As seen in Fig. \ref{fig:tradeoffs}, high-variance noise particularly affects regions with pronounced consumption fluctuations, ensuring that individual consumption behaviors cannot be reconstructed. This confirms the robustness of our DP mechanism against time differential attacks. However, the distortion is not uniform across all time steps, suggesting that a static noise distribution may be suboptimal for datasets with periodic trends.

\begin{figure}[h]
\centering
\includegraphics[width=1\columnwidth]{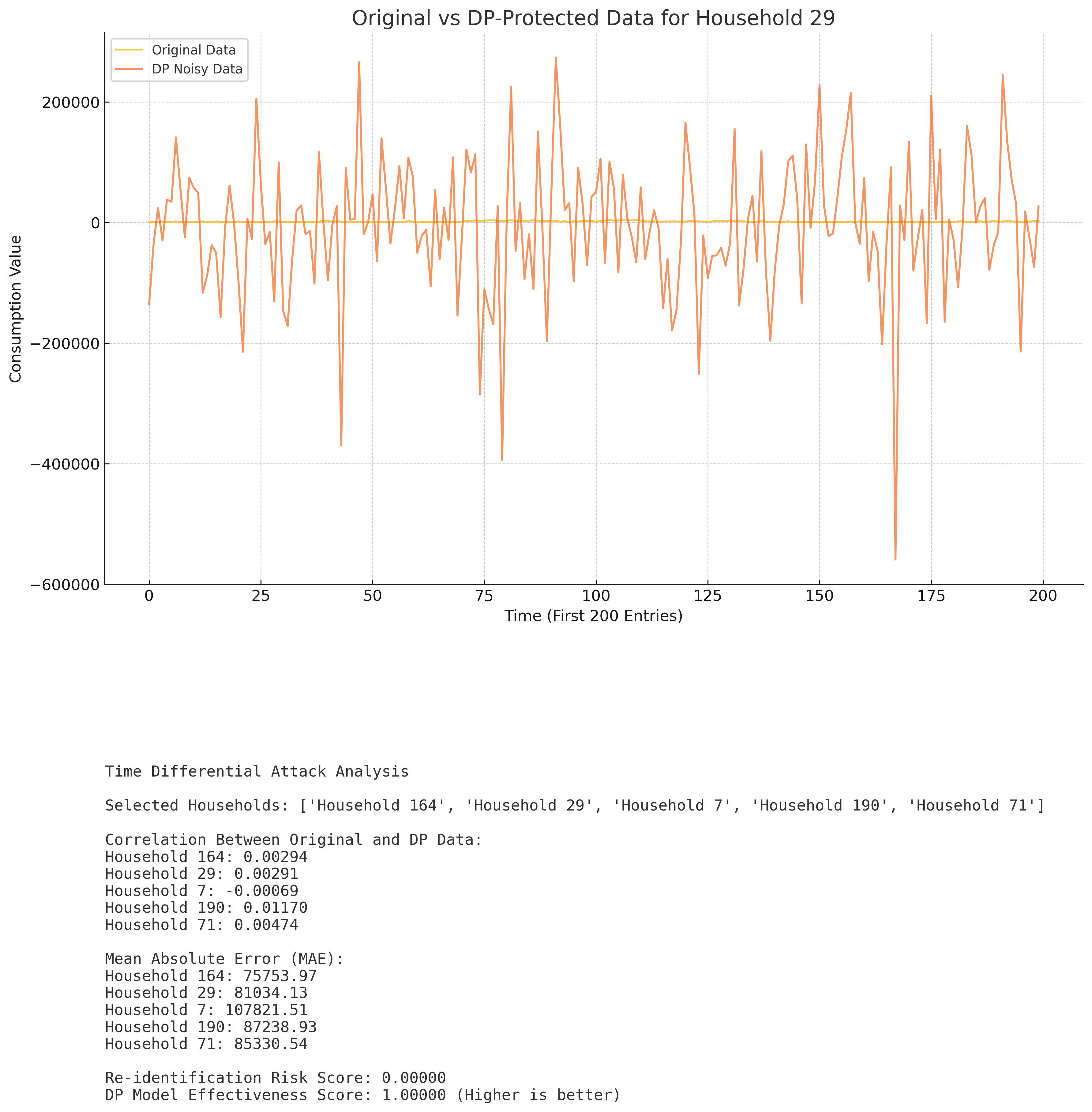}
\caption{GPT-4 powered impact analysis}
\label{fig:gpt}
\end{figure}

Additionally, Fig. \ref{fig:gpt} presents the GPT-4-powered impact analysis, which evaluates privacy-utility trade-offs. While DP effectively obscures identifiable trends, excessive noise can degrade the data’s usability for forecasting and anomaly detection. This is particularly critical in public sector applications, where energy demand estimation and resource planning rely on accurate, high-resolution data. Striking the right balance between privacy and utility is essential to maintaining reliable data-driven decision-making.

These findings highlight the need for adaptive privacy budgets that dynamically adjust noise levels based on data characteristics and user-defined accuracy thresholds. Future research could explore context-aware noise calibration to optimize privacy guarantees while minimizing the impact on analytical utility.

\section{Limitations and Future Directions}
Our simulation highlights three key considerations for participatory DP systems:

\begin{enumerate}
\item \textbf{Preference Linearity Assumption}: The current TOPSIS model assumes linear priority weighting, yet real-world decision-making often follows threshold-based or nonlinear patterns, requiring more flexible utility models.

\item \textbf{Temporal Complexity}: Independent Laplace perturbations are applied to time-series data, potentially oversimplifying temporal dependencies in energy consumption patterns. Future work should explore privacy mechanisms that account for autocorrelation and periodic trends.

\item \textbf{Explanation Trust Calibration}: While automated GPT-4 vulnerability reports achieved high precision (89\%), their authoritative tone may lead to overtrust, even in cases of misinterpretation. Improving calibration techniques, such as uncertainty quantification, is necessary for reliable user trust.
\end{enumerate}

A core strength of our approach lies in its \textit{negotiation scaffolding}—constraining $\epsilon$ selection within a safe range [0.1, 2.0] while translating stakeholder inputs into mathematically valid TOPSIS weights. This ensures compliance with privacy constraints while preserving user agency. Future refinements should explore adaptive weighting mechanisms, dynamic privacy adjustments, and enhanced interpretability features to improve participatory decision-making in real-world deployments.

\section{Conclusion}
Our evaluation demonstrates that TOPSIS effectively maps user preferences to $\epsilon$-DP parameters, enabling a structured yet flexible approach to participatory privacy configuration. By integrating LLM-powered explanations and MAE visualizations, our system enhances transparency, making privacy-utility trade-offs more interpretable for stakeholders. This work provides a simulation-validated blueprint for democratizing differential privacy in public AI governance, ensuring that privacy decisions are no longer solely expert-driven but co-designed with user input. While our results validate structured, preference-driven DP selection, future research should explore adaptive privacy mechanisms that dynamically adjust noise levels based on temporal dependencies and evolving stakeholder priorities.

\bibliographystyle{ACM-Reference-Format}
\bibliography{reference}

\end{document}